\documentclass{aastex6}

\usepackage[utf8x]{inputenc}
\usepackage{natbib}
\usepackage{graphicx}
\usepackage{multirow}
\usepackage{mathrsfs,amssymb}
\usepackage{amsmath}

\newcommand       \AU           {\,{\rm AU}}

\newcommand       \ms           {\,{\rm m/s}}
\newcommand       \kms           {\,{\rm km/s}}

\newcommand       \pc           {\,{\rm pc}}

\newcommand       \Myr          {\,{\rm Myr }}

\newcommand       \Msun         {\,{M_\odot}}

\newcommand       \Lsun         {\,{L_\odot}}

\newcommand{\jj}[2]{\mbox{$J = #1\rightarrow#2$}}
\newcommand{\msun}{\mbox{M$_\odot$}}
\newcommand{\coo}{$^{13}$CO}
\newcommand{\cooo}{C$^{18}$O}

\begin{document}

\title{The Effects of Protostellar Disk Turbulence on CO Emission
Lines: A Comparison Study of Disks with Constant CO Abundance
vs.\ Chemically Evolving Disks}

\author{Mo Yu\altaffilmark{1}}
\author{Neal J. Evans II\altaffilmark{1, 2}}
\author{ Sarah E. Dodson-Robinson\altaffilmark{3}}
\author{Karen Willacy\altaffilmark{4}}
\author{ Neal J. Turner\altaffilmark{4}}

\altaffiltext{1}{Astronomy Department, University of Texas, 2515 Speedway, Stop C1400, Austin, TX 78712, USA}
\altaffiltext{2}{Korea Astronomy and Space Science Institute, 776, Daedeokdae-ro, Yuseong-gu, Daejeon, 34055, Korea}
\altaffiltext{3}{University of Delaware, Department of Physics and Astronomy, 217 Sharp Lab, Newark, DE 19716  }
\altaffiltext{4}{Mail Stop 169-506, Jet Propulsion Laboratory, California Institute of Technology, 4800 Oak Grove Drive, Pasadena, CA 91109}

\altaffiltext{1}{Astronomy Department, University of Texas, 1 University Station C1400, Austin, TX 78712, USA}
\altaffiltext{2}{University of Delaware, Department of Physics and Astronomy, 217 Sharp Lab, Newark, DE 19716  }
\altaffiltext{3}{Mail Stop 169-506, Jet Propulsion Laboratory, California Institute of Technology, 4800 Oak Grove Drive, Pasadena, CA 91109}

\begin{abstract}

Turbulence is the leading candidate for angular momentum transport in
protoplanetary disks and therefore influences disk lifetimes and planet
formation timescales. However, the turbulent properties of
protoplanetary disks are poorly constrained observationally.
Recent studies have found turbulent speeds smaller than what fully-developed MRI would
produce (Flaherty et al.\ 2015, 2017). However, existing studies assumed a constant CO/H2
ratio of $10^{-4}$ in locations where CO is not frozen-out or photo-dissociated. Our previous studies of evolving disk chemistry indicate that CO is depleted by incorporation into complex organic molecules well inside the freeze-out radius of CO. We consider the effects of this chemical depletion on measurements of turbulence. Simon et al.\ (2015) suggested that the ratio of the peak line flux to the flux at line center of the CO J=3-2 transition is a reasonable diagnostic of turbulence, so we
focus on that metric, while adding some analysis of the more complex effects on spatial distribution.
We simulate the emission lines of CO based on chemical
evolution models presented in \citet{Yu_2016_COchem}, and find that the
peak-to-trough ratio changes as a function of time as CO is destroyed.
Specifically, a CO-depleted disk with high turbulent velocity mimics the
peak-to-trough ratios of a non-CO-depleted disk with lower
turbulent velocity. We suggest that disk observers and modelers take
into account the possibility of CO depletion when using line
peak-to-trough ratios to constrain the degree of turbulence in disks.
Assuming { that CO/H$_2 = 10^{-4}$ at all disk radii} can
lead to underestimates of turbulent speeds in the disk by at least $0.2
\kms$.

\end{abstract}

\section{Introduction}


An angular momentum transfer mechanism is essential for the evolution of disks, the growth of stars, and the formation of planets. 
Throughout much of a T-Tauri disk, the combination of
ionizing radiation and Keplerian shear should trigger the
magnetorotational instability \citep[MRI;][]{Balbus_Hawley_1991,
balbus98, hawley01, Fromang_Nelson_2006, salmeron08}, 
which drives accretion rates of
order $\sim 10^{-9} M_{\odot}$~yr$^{-1}$ \citep{Bai_2011, Landry_2013,
Simon_2013}.
An alternative to MRI turbulence comes from recent simulations showing that magnetic winds
may drive angular momentum transfer in protoplanetary disks
\citep{Bai_Wind1_2013, Bai_Wind2_2013, lesur14, Gressel_2015,
bai16}. In the magnetocentrifugal wind model of
\citet{Gressel_2015}, the disk remains laminar between 1 and 5~AU, with no significant turbulence. Since the 
gas velocity field controls both the sticking efficiency of colliding dust grains
\citep[e.g.][]{dominik97,blum08,birnstiel10,zsom11} and the mass loss
due to erosion or fragmentation in pebble collisions
\citep[e.g.][]{Brauer_2008, Birnstiel_fragmentation_2009, guttler10,
kothe10, beitz11}, empirical measurements of turbulent speeds in disks
would be extremely useful in developing theories of solid accretion, as well as understanding disk evolution.

Unfortunately, turbulent velocity profiles in disks are not { always}
well constrained.  \citet{Flaherty_2015_turbulent_observed} observed CO
emission lines in the HD~163296 disk with ALMA. The observations were
interpreted as limiting turbulent speeds to levels far below what
fully-developed MRI would produce, and not strong enough to explain the
star's high accretion rate of $5 \times 10^{-7} M_{\odot}$~yr$^{-1}$
\citep{mendigutia13}. Yet most turbulent speed measurements are
model-dependent, and the HD~163296 disk may have complex CO chemistry
that was not included in the \citet{Flaherty_2015_turbulent_observed}
model. In a direct (non model-based) analysis of new ALMA CO, CN, and CS
observations of TW~Hydrae,
\citet{Teague_2016} argue that almost all literature
measurements of turbulent speeds---whether direct or
model-based---should be interpreted as upper limits due to ALMA's 7\%
flux calibration accuracy. According to \citet{Teague_2016}, firm
detections of turbulent motion require a ratio of turbulent speed to
sound speed of $v_{\rm turb} / c_s \ga 0.1$. Here we introduce another
cautionary note by demonstrating how complexities in the spatial
distribution of CO abundance can affect line shapes and complicate
turbulent speed measurements.

Rotational emission lines from disks naturally form
double-peaked profiles with a ``trough'' at the systemic velocity.
Based on models of MRI-active disks, \citet{Simon_2015} found the ratio
of the peak line flux to the flux at the line center (peak-to-trough
ratio) to be a reasonable diagnostic of the MRI turbulence. However, 
\citet{Simon_2015} and \citet{Flaherty_2015_turbulent_observed}---who modeled the entire ALMA data cube---assumed a constant CO/H$_2$ ratio of $10^{-4}$ in locations where CO is
not frozen-out or photo-dissociated. Our chemical evolution models (\citet{Yu_2016_COchem}, hereafter Paper 1) indicate that the CO
abundance is a complex function of both radius and time, and CO
abundance gradients may affect the peak-to-trough ratio of CO emission
lines. This paper examines the effects of CO depletion on peak-to-trough
ratios and turbulent speed measurements. We summarize the results of the thermal-chemical models in section \ref{sec:diskmodels}, 
and describe the molecular line radiative transfer models in section \ref{sec:limemodels}. We discuss the implications of CO chemical depletion for line profiles and turbulent speed measurements in section \ref{sec:profile}. Finally, we
examine the changes in peak-to-trough ratio as the disk evolves in section \ref{sec:time_evo_p2t}.

\section{Thermal-chemical models}
\label{sec:diskmodels}

We adopt the thermo-chemical model from Paper 1 and \citet{Yu_2017_mass} 
(hereafter Paper 2) as the basis for this study. Paper 1 presented the chemical 
evolution of a $0.015\Msun$ disk around a Solar-type star for $3$~Myr, and Paper 2 introduced the model of a $0.03\Msun$ disk, following the same modeling procedures as used for the $0.015\Msun$ disk model. We summarize the key results of the two models in this section and refer readers to our previous papers for details.

 \citet{Landry_2013} presented the mass distributions and temperatures,
considering only accretion heating, for both model disks, and we further calculated the stellar contribution to the disk heating with the dust radiative transfer code RADMC \footnote{http://www.ita.uni-heidelberg.de/~dullemond/software/radmc-3d/; developed by C. Dullemond} in Paper 1 and Paper 2. We then ran the chemical reaction network locally at each independent (r, z) grid point for $3$ Myr, following the viscous evolution of the disk and the evolution of the central star along the Hayashi track. No mixing is included in the model; we assume the chemical reaction timescale to be much shorter than the viscous timescale, which is true for freezeout, desorption, and grain-surface reactions, but which may fail for gas-phase reactions. Each disk gridpoint starts with gas and ice abundances resulting from a 1 Myr simulation of the chemical evolution of a parent molecular cloud; as a result, a substantial fraction of the carbon is tied up in CO$_2$ and other ices at the start of disk evolution.

The chemical evolution models include C, H, O, N based on the UMIST database
RATE06 \citep{Woodall_UMIST_2007}. \citet{Woods_Willacy_2009} extended
the network to include C isotopes, and we included both C and O isotopes
in Paper I. The chemical models follow the chemistry of $588$ species,
$414$ gas-phase and $174$ ices for $3\Myr$ from the beginning of the
T-Tauri phase. The reaction network contains gas-phase reactions,
grain-surface reactions, freezeout, thermal desorption, and reactions
triggered by UV, X-rays, and cosmic rays, such as isotope-selective
photodissociation. 

\begin{figure*}[ht]
\centering
\begin{tabular}{@{}cc@{}}
\includegraphics[width=0.45\textwidth]{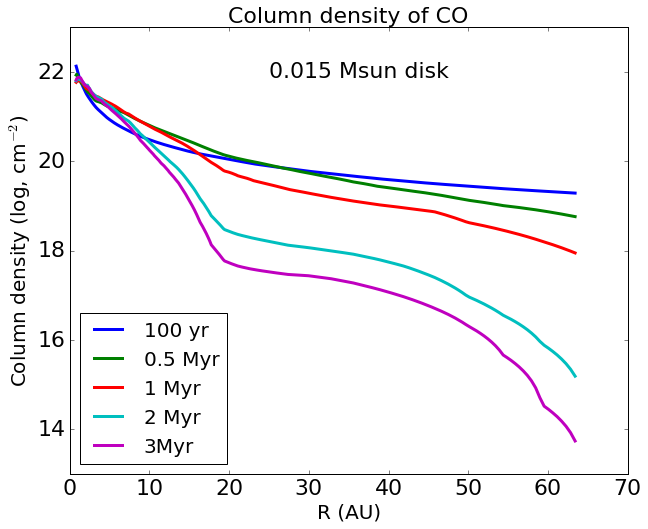} &
\includegraphics[width=0.45\textwidth]{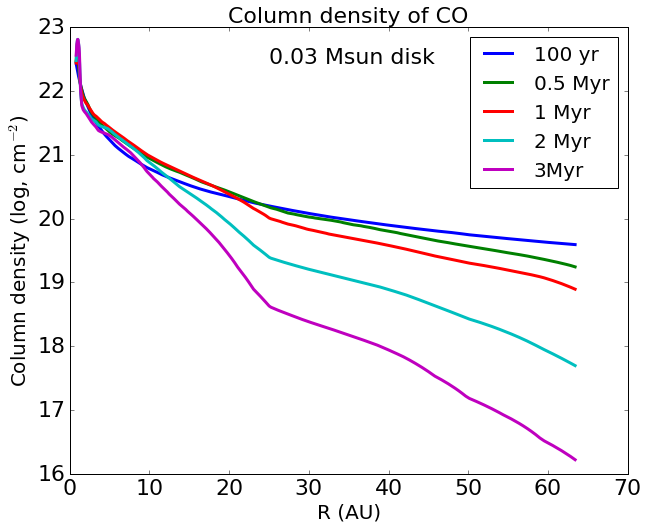} \\
\includegraphics[width=0.45\textwidth]{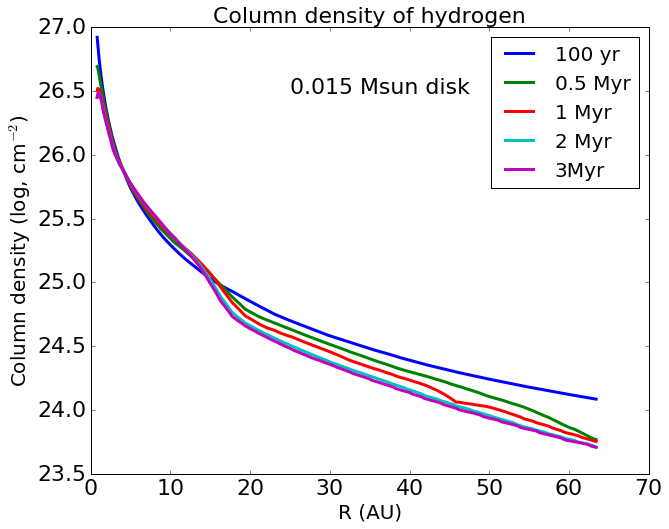} &
\includegraphics[width=0.45\textwidth]{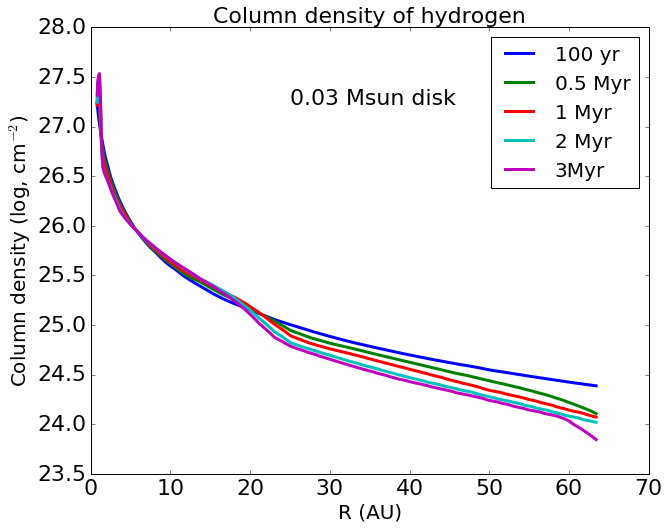} \\
 \end{tabular}
\caption{Column densities of CO (upper panels) and H$_{2}$ (lower panels) as a function of disk radius (R) in different stages of the disk evolution. Results for the $0.015 \Msun$ model are shown on the left and the results for the $0.03 \Msun$ model are on the right. }
\label{fig: coldens}
\end{figure*}


The luminosity of the central star ranges from $12 \Lsun$ to about $0.8 \Lsun$ over the $3$Myr evolution. Due to efficient heating from the central star, CO does not freeze in our modeled region---the inner 70 AU of the disk---at any time in the $3$~Myr of evolution. However, CO is depleted beyond $20$~AU from the central star due to the formation of complex organic molecules, a process that we call
chemical depletion. The CO chemical depletion is driven by ionization of helium from X-rays and cosmic rays and happens over a million-year time scale. As a result, the CO abundance changes both with location in the disk and with time. We show the column density as a function of disk radius for CO and H$_{2}$ at different stages of the disk evolution for both the $0.015 \Msun$ and the $0.03 \Msun$ disks in Fig. \ref{fig: coldens}. The column density of CO drops significantly beyond $20$ AU in both disks after roughly a million years of disk evolution, while the radial profile of the H$_2$ column density changes very little over time. The small change in H$_{2}$ column density is caused mostly by the pileup of material in the MRI dead zone (see paper 1 for details), while the change in CO column density is caused by the gradual chemical depletion of CO, with the liberated carbon atom forming complex species which then freeze, sequestering carbon in organic ices. The thermal-chemical model results are qualitatively similar for the disks of two masses (Papers 1 and 2). The chemical depletion timescale is slightly longer for the $0.03 \Msun$ disk because the higher column density decreases the ionization fraction---and thus the abundance of ionized helium---in the midplane. To complicate matters, while the outer disk is losing its CO, the abundance of CO simultaneously {\it increases} with time at small radii, as CO$_2$ ice is converted to CO gas. { Figure \ref{fig: coldens} suggests that CO chemical depletion may affect molecular line emission by decreasing the emitting area as the disk evolves. As the CO abundance drops in the outer disk, the disk location where the optical depth drops below unity moves inward.}

The net result of our chemical models is that CO becomes severely depleted well inside the CO freeze-out radius in disks with masses above the minimum needed to form planetary systems. Similar effects have been seen in other chemical evolution calculations (\citealt{1999ApJ...519..705A, Furuya_carbon_2014, 2014A&A...563A..33W, Bergin14}). A discussion of the similarities and differences between our models and other work can be found in paper 2. 
\section{Line radiative transfer models}
\label{sec:limemodels}

After modeling disk thermal and chemical structures as functions of time, we build molecular line radiative transfer models to simulate observational properties of the disk. We use the publicly available code LIME \citep[LIne Modeling Engine]{Brinch_Hogerheijde_LIME_2010}, and adopt the energy levels and collision rates from the Leiden Atomic and Molecular Database (LAMDA)\footnote{http://home.strw.leidenuniv.nl/~moldata/}. As a first order approximation, we do not consider the hyperfine splitting in C$^{17}$O emission. 

We model emission from within $70\AU$ of the central star, which
corresponds to a $1\arcsec$ beam diameter for an assumed distance of
$140\pc$ from the Sun. We consider line broadening due to Keplerian
rotation, thermal velocity and micro-turbulence. Thermal velocities are
calculated assuming a Maxwell-Boltzmann speed distribution based on the
disk's temperature structure from Paper 1. We assume an isotropic
Maxwell-Boltzmann speed distribution with RMS of $100\ms$ everywhere in
the disk for the micro-turbulence in the fiducial models, and discuss
effects of varying both the RMS turbulent speed and its radial and
vertical profile in later sections. To compute the level populations, we
set a minimum scale of $0.07\AU$ to guarantee sub-pixel sampling of both
Keplerian speeds and CO abundance gradients. We first generate the
synthetic datacube of intensity as a function of $x$, $y$, and velocity
for a disk around a $0.95 \Msun$ star at $30\degr$ inclination.
In velocity space, the spectra have
$300$ channels of $125\ms$ resolution. At any specific velocity, the
synthetic image contains $600 \times 600$ pixels of $0.003\arcsec \times
0.003\arcsec$ in size. Finally, we generate the synthetic spectra
presented here by integrating each velocity component over a square with
$1.2\arcsec$ sides ($400 \times 400$ pixels), larger than the angular
size of the disk. The pixels not covered by the disk contribute no flux
and are included simply for ease of integration---here we assume that
the sky background contains negligible flux compared with the disk at
all wavelengths.

Our current models assume that the gas temperature is the same as the
dust temperature. As we demonstrated in the Appendix of Paper 2, this is
a valid assumption for $\jj32$ and $\jj21$. However, the difference
between the gas and dust temperature would need to be considered in
order to use our models to fit high-J spectral lines, which are
dominated by emission from the disk surface. For the rest of the paper,
we focus our line profile discussion on the $\jj32$ and $\jj21$
transitions, which are the most commonly observed. We assume LTE for
energy level populations, an assumption that we have justified in the
Appendix of Paper 2.


\section{Effects of CO depletion and turbulent velocities on CO line profiles}
\label{sec:profile}

We show the time evolution of the CO $\jj32$ line from our
fiducial disk of mass 0.15~\msun\
on the left side of Figure \ref{fig: line_evo}. Simulated emission lines
from LIME are plotted in the upper panel of each plot, and the profiles
normalized to the peak intensities are shown in the lower panels. The
total intensities decline with time dramatically, { the wings become
broader in the normalized line profile}, and the relative contribution
from the line center decreases over time. The line's broadening over
time occurs because CO depletion happens primarily in the outer part of
the disk where Keplerian velocities are small, so the fraction of
radiation from the high-velocity line wings increases with time.
Moreover, new CO forms from CH$_{3}$ and CO$_{2}$ at small radii,
further increasing the contribution of the line wings. The bottom-left
panel of Figure \ref{fig: line_evo} shows that the peak-to-trough ratio
of CO~$\jj32$ must increase as the fiducial disk evolves.

 \subsection{Effects of the chemical depletion of CO}
 \label{subsec: profiles_co_depletion}

\begin{figure*}[ht]
\centering
\begin{tabular}{@{}cc@{}}
 \includegraphics[width=0.4\textwidth]{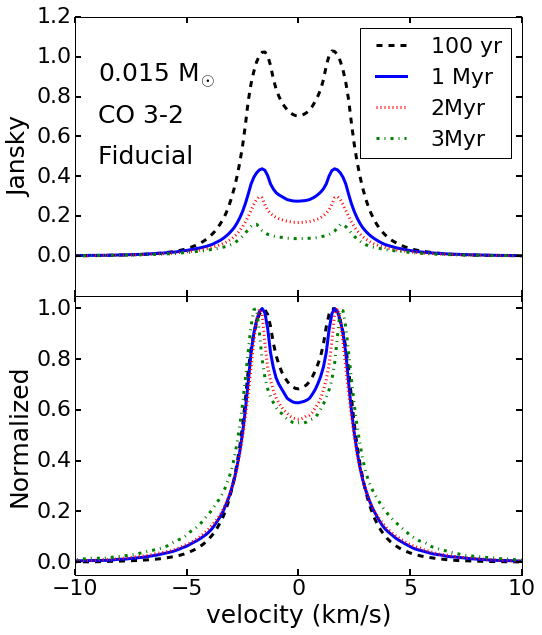} &
  \includegraphics[width=0.4\textwidth]{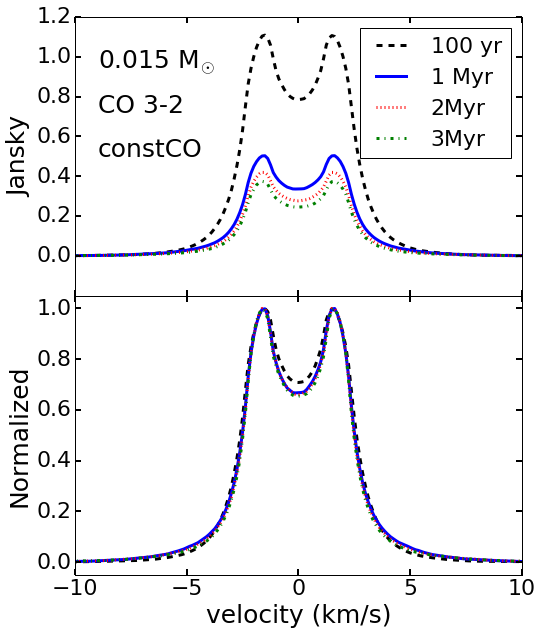} \\
 \end{tabular}
  \caption{Time evolution of the CO $\jj32$ line. Results from the fiducial model with $0.015 \Msun$ are on the left, and results from the constant CO model of the same mass are on the right. The top panels show the simulated lines from LIME, and the lower panels show the line profiles normalized to the peak intensity of each line. In both models, emission becomes weaker over time as the disk cools. The {\it relative} contribution from the line center also decreases over time in the fiducial model with CO chemical depletion, while normalized line profiles in the constant CO model remain the same for the last $2$ Myr of the disk evolution.}
 \label{fig: line_evo}
\end{figure*}

To isolate the effect of chemical depletion of CO on emission line
profiles from effects caused by the evolution of disk density and
temperature, we introduce models with constant CO abundance throughout
the disk---assuming all available carbon is in the form of CO gas. The
abundance of CO normalized to the total proton number density is $7.21
\times 10^{-5}$. All the other properties of the constant CO models are
the same as in the fiducial model. The time evolution of the CO $\jj32$
line for the constant CO model are shown on the right of Figure
\ref{fig: line_evo} for comparison. The models with constant CO predict
a much smaller decrease in integrated line intensity with time than do
the fiducial models. As most easily seen in the normalized profiles
(lower panels), the peak-to-trough ratio changes much more in the
fiducial models, which account for chemical depletion, than in the
constant CO models.

{ Paper 2 describes how CO chemical depletion biases disk-mass
measurements, leading to severe underestimates when the interstellar
CO/H$_2$ ratio of $10^{-4}$ is assumed. In \S 6 of Paper 2, we discuss
three strategies for diagnosing CO depletion: (1) measuring the
C$^{18}$O/$^{13}$CO integrated intensity ratio, which drops sharply as
the disk evolves; (2) comparing line profiles of $^{13}$CO and either
C$^{18}$O or C$^{17}$O, as the decrease in emitting area as a function
of time preferentially broadens profiles of optically thin lines; or (3)
comparing spatially resolved maps from an abundant vs.\ a rare
isotopologue (e.g.\ CO and C$^{18}$O), as the emission from the rare
molecule will be truncated at the CO depletion front at $\sim 20$~AU.
Although CO depletion greatly decreases total line flux, the flux is
also a complex function of disk mass and temperature, which is why we do
not consider flux from a single emission line an adequate diagnostic of
chemical depletion.  Given that CO chemical depletion affects
peak-to-trough ratios, we recommend using the procedures described in
paper 2 to diagnose the degree of CO depletion when interpreting either
peak-to-trough ratios or full line profiles. In \S
\ref{sec:time_evo_p2t} we further explore how peak-to-trough ratios
change as the outer disk loses its CO.}

\subsection{Effects of varying RMS turbulent speed and radial/vertical speed profile}
\label{subsec: profiles_turbulent}


Next, we want to isolate the effect of turbulent velocities on line
profiles. All models shown in Figure \ref{fig: line_evo} include a
microturbulent velocity field with RMS of $0.1\kms$. To distinguish the
effects of CO depletion from the effects of RMS turbulent speed on CO
line profiles and therefore the peak-to-trough ratios, we simulate CO
and C$^{18}$O $\jj32$ emission for a $2$~Myr fiducial disk with a range
of RMS micro-turbulent velocities.

\begin{figure*}[ht]
\centering
\begin{tabular}{@{}cc@{}}
\includegraphics[width=0.45\textwidth]{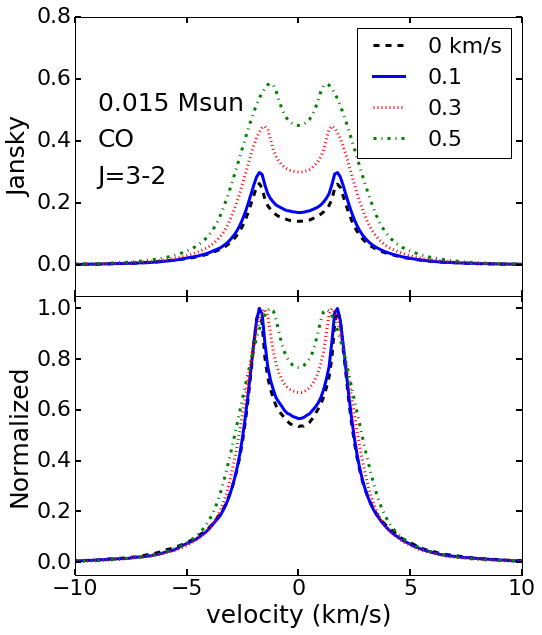} &
\includegraphics[width=0.45\textwidth]{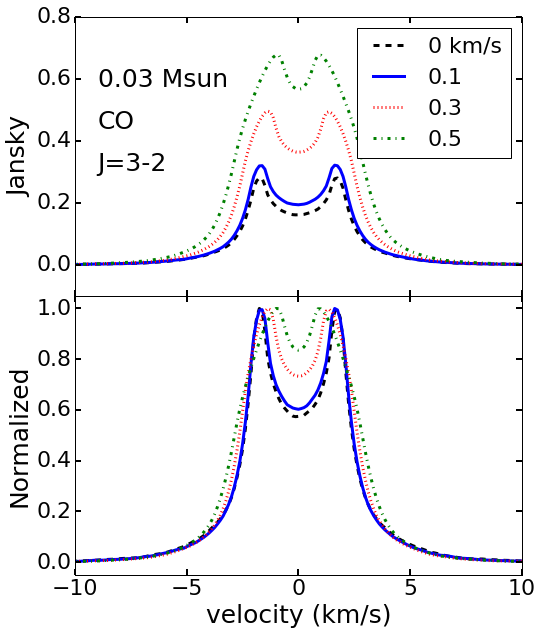} \\
 \end{tabular}
\caption{The line profiles of CO for a 2 Myr disk with various micro-turbulent velocities. The $0.015 \Msun$ model is on the left and the $0.03 \Msun$ model is on the right. In both disks, the increase of micro-turbulent velocity increases the relative contribution from the line center, resulting in a decreasing peak-to-trough ratio.}
\label{fig: line_bexp}
\end{figure*}

We show the CO line profiles from both the $0.015 M_{\odot}$ and the
$0.03 M_{\odot}$ fiducial disk models at the 2~Myr time snapshot, with
varying micro-turbulent velocities, in Fig \ref{fig: line_bexp}. In both
disks, the increase of micro-turbulent velocity leads to increased
contributions from the line center, resulting in smaller peak-to-trough
ratios. Similar effects were found by \citet{Simon_2015}---for a single
time snapshot of the disk model, the peak-to-trough ratio decreases with
increasing microturbulent velocity. Integrated intensity increases with
RMS turbulent speed due to the opacity decrease from spreading the
absorbers in velocity space.

{ While CO depletion strongly increases the peak-to-trough ratio but
has only a small effect on the velocity spacing between the line profile
peaks (see Figure \ref{fig: line_evo}), decreasing the RMS turbulent
speed simultaneously increases the peak-to-trough ratio and
widens the $\Delta v$ between the line profile peaks
(Figure \ref{fig: line_bexp}). Since both CO depletion and strong
turbulence broaden the emission line, the combination of a high
peak-to-trough ratio (which, for interstellar CO/H$_2 = 10^{-4}$, would
indicate weak turbulence) {\it and} a broad normalized line profile
probably indicates some degree of CO depletion.  Analyzing the entire
line profile rather than the peak-to-trough ratio, as did
\citet{hughes11} and \citet{guilloteau12}, would improve estimates of turbulent
speed.}

\begin{figure*}[ht]
\centering
\begin{tabular}{@{}cc@{}}
\includegraphics[width=0.45\textwidth]{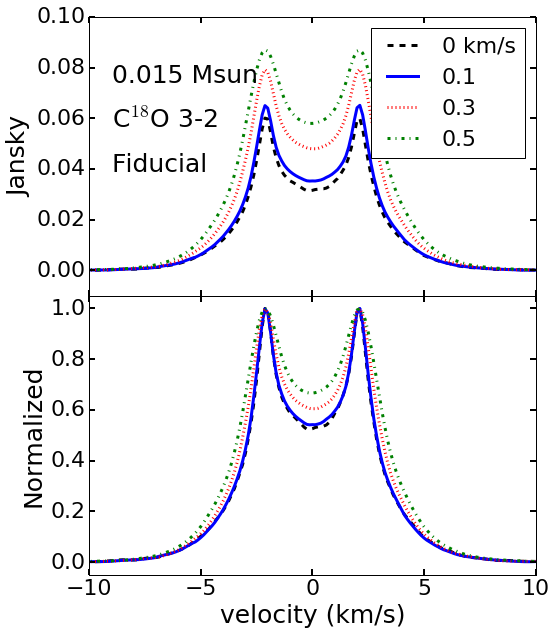} &
\includegraphics[width=0.45\textwidth]{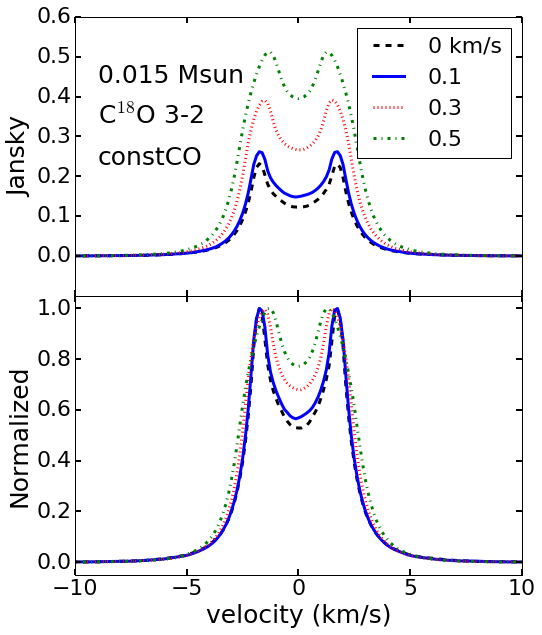} \\
 \end{tabular}
\caption{Effect of turbulent velocity changes on the C$^{18}$O
$\jj32$ line emitted by the $0.015 \Msun$ disk at age 2 Myr. Left: fiducial model with CO depeltion. Right: constant CO model.}
\label{fig: c18o_vturb}
\end{figure*}

{ Figure \ref{fig: c18o_vturb} shows \cooo\ line profiles for the
$0.015 \msun$ disk at 2~Myr, with the fiducial, chemically evolving
model on the left and the constant-CO model on the right. With
\cooo\, a
high velocity spread between the line profile peaks indicates chemical
depletion: we find a peak-to-peak velocity spread of $\Delta v = 4.15 \kms$ 
in the fiducial model with
$v_{\rm tur} = 0.1 \kms$ vs.\ $\Delta v = 3.37 \kms$ for the constant-CO
model with $v_{\rm tur} = 0.1 \kms$. Like CO, \cooo\ shows the same
decrease in peak-to-trough ratio with increasing $v_{\rm tur}$. However,
unlike CO, changes in $v_{\rm tur}$ do not affect the velocity spacing
between the line peaks for the fiducial model (though they affect the
peak spacing for the constant-CO model). Furthermore, the peak-to-trough
ratios from the fiducial and constant-CO models are not very different
when the turbulence is weak (for example, with $v_{\rm tur} = 0.1 \kms$,
the fiducial model has P/T~=~1.84 while the constant-CO model has
P/T~=~1.76).  Given that most observational studies suggest weak
turbulence in T-Tauri and Herbig Ae/Be disks
\citep[e.g.][]{hughes11,guilloteau12,Flaherty_2015_turbulent_observed,Teague_2016,flaherty17},
\cooo\ may be a more robust turbulence indicator than CO if the degree of
chemical depletion is not known. As in Paper 2, we recommend comparing
\cooo\ line profiles with those of CO or \coo\ to diagnose chemical
depletion.}

\begin{figure*}[ht]
\centering
\begin{tabular}{@{}cc@{}}
\includegraphics[width=0.45\textwidth]{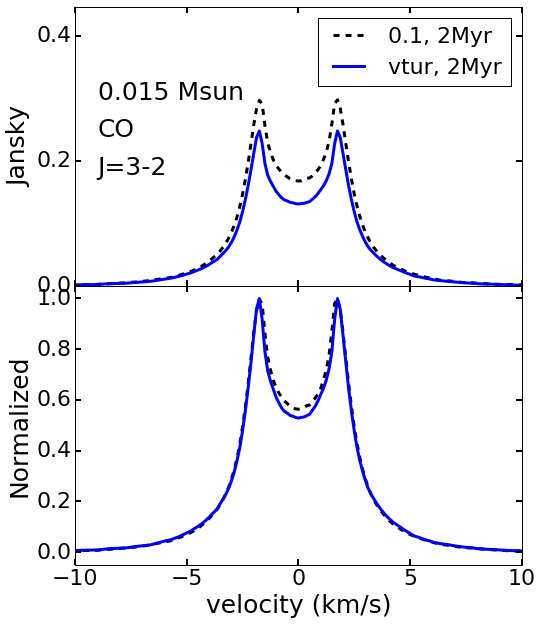} &
\includegraphics[width=0.45\textwidth]{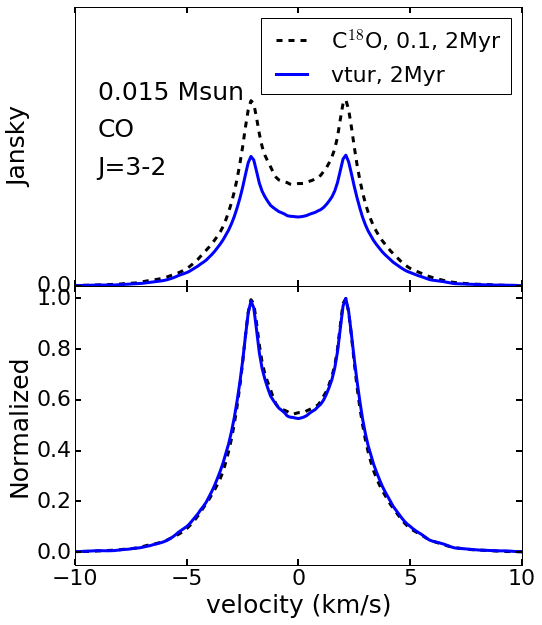}
 \end{tabular}
\caption{Line profiles of CO emission from a 2~Myr, $0.015 M_{\odot}$ disk with the turbulent speed profile from the MRI model vs.\ a constant RMS speed of $0.1\kms$. { Left:} CO~$\jj32$; { Right:} C$^{18}$O~$\jj32$. Neither molecule produces line profiles that probe the vertical turbulent speed profile.}
\label{fig:turbulence_mri}
\end{figure*}

Besides the magnitude of the RMS turbulent speed, the radial and
vertical dependence of the turbulence speed can also affect the line
profile and the peak-to-trough ratio. Landry et al.\ provide the spatial
distribution in $(r,z)$ of the RMS turbulent speed in each grid cell as
a function of time in the form of $\alpha$, where the 1D turbulent
velocity is $v_{tur} = \sqrt{3} \alpha \times C_{s}$, with $C_{s}$ as
the local sound speed. We include the turbulent velocity $(r,z)$ profile
for the $0.015\msun$ disk at $2$ Myr to see the effect of incorporating
non-uniform turbulent velocities in Figure \ref{fig:turbulence_mri}. The
\citet{Landry_2013} model predicts a nearly quiescent midplane dead zone
from 2-20 AU, with low turbulent speeds near the disk surface that
generate an accretion rate onto the star of $\sim 10^{-9}
\Msun$~yr$^{-1}$, { topped by a dead atmosphere with $v_{\rm tur} <
1 \ms$ due to ambipolar diffusion (see their Figure 1). The
non-turbulent atmosphere provides a slightly higher peak-to-trough ratio
in CO~$\jj32$ than seen in the model with $100 \ms$ RMS turbulent speed
throughout the disk, though extremely high S/N observations would be
required to tell the difference between the two models.}

{ Although C$^{18}$O peak-to-trough ratios are less sensitive to
chemical depletion than CO peak-to-trough ratios, \cooo\ does not perform
well as a diagnostic of non-uniform turbulent speed for our model disk.
Figure \ref{fig:turbulence_mri} shows that the normalized line profiles
from the Landry et al.\ model and the disk with radially and vertically
constant $100 \ms$ turbulence are indistinguishable. This is because
most of the C$^{18}$O emission emerges from the slightly active layer
where the turbulent speeds are $\sim 40 \ms$. Given that most of the CO
gas (including all isotopologues) is concentrated in the dead zone,
emission from near the disk midplane should produce line profiles
similar to those from the completely laminar, $0 \kms$ turbulent-speed
models shown in Figure \ref{fig: line_bexp}. We find, therefore, that
C$^{18}$O emission does not trace the midplane dead zone. The close
match between line profiles from the Landry et al.\ disk and the disk
with constant $100 \ms$ RMS turbulent speed comes from the fact that gas
at both disks' $\tau \sim 1$ surfaces has a similar speed distribution.}

\section{Time evolution of the peak-to-trough ratio}
\label{sec:time_evo_p2t}

Figure \ref{fig: line_evo} shows that disks with the same mass
distribution and ionization environment will have different CO emission
line profiles if observed at different ages. We calculate peak-to-trough
ratios as a function of time and plot those of the J=$3-2$ lines in
Figure \ref{fig: peak2trough}. In the fiducial model, the CO
peak-to-trough ratio increases from about 1.45 to 1.85 as the disk
evolves. The change over time is comparable to the increase of
peak-to-trough ratio caused by the decrease of turbulent velocities by a
factor of 2 to 3, as found in both \citet{Simon_2015} and Section
\ref{subsec: profiles_turbulent}. The growth of the high-velocity line
wings induced by chemical evolution can masquerade as a decrease in RMS
turbulent speed with time if peak-to-trough ratio is taken as an
indicator of {\it only} thermal and turbulent speed.

\subsection{Assuming {\rm CO/H}$_2 = 10^{-4}$ leads to underestimates of turbulent speed}

\begin{figure*}[ht]
\centering
\begin{tabular}{@{}cc@{}}
\includegraphics[width=0.45\textwidth]{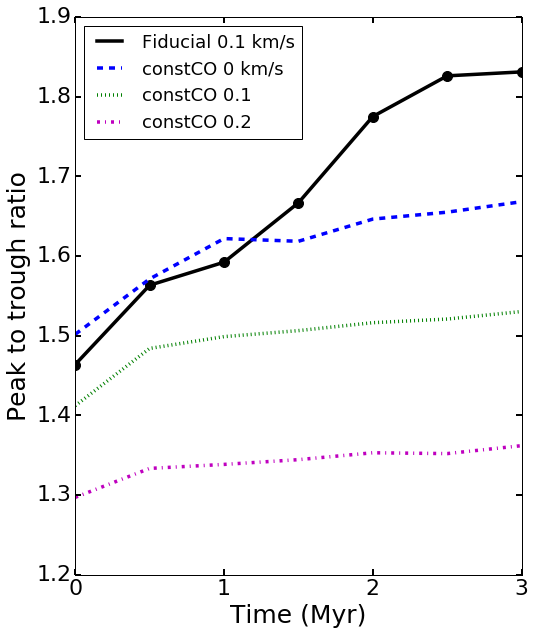} &
\includegraphics[width=0.45\textwidth]{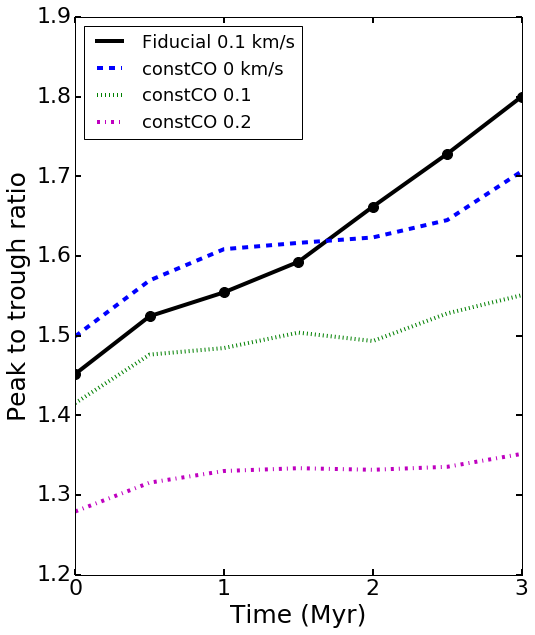} \\
 \end{tabular}
\caption{The time evolution of the peak-to-trough ratios of the CO $\jj32$ line. The $0.015 \Msun$ disk is plotted on the left and the $0.03 \Msun$ model is on the right. The black lines are the fiducial model (with CO chemical depletion, and turbulent velocity of $0.1\kms$). For comparison, we plot the peak-to-trough ratio predicted assuming a constant CO abundance with different turbulence velocities in blue, green and magenta. }
\label{fig: peak2trough}
\end{figure*}

Would an observer misinterpret the high peak-to-trough ratios predicted
in our models as signatures of low turbulent velocities? To answer this
question, we calculate both constant-CO models and fiducial models with
chemical depletion, incorporating a range of turbulent velocities. We
then consider how observers would interpret peak-to-trough ratios with
no prior knowledge of disk chemical composition. The evolutionary tracks
of the peak-to-trough ratio assuming no CO depletion with various
turbulent velocities are shown alongside those of the fiducial model in
Fig.\ \ref{fig: peak2trough}.

The peak-to-trough ratio increases slowly over time even without the
effect of CO chemical depletion. During the first million years of
evolution, this is due to significant disk cooling as the star dims
along the Hayashi track. { The decrease of emitting area during CO
depletion and the disk cooling combine to increase the peak-to-trough
ratio later in the evolution, given that we see the peak-to-trough ratio
continue to increase after 1~Myr in our fiducial model but
begin to level off after 1~Myr in the constant-CO model.} While the fiducial model mimics the peak-to-trough ratios of
models with lower turbulent velocities in the first million years of
evolution, it produces peak-to-trough ratios higher than what could be
explained by the grid of constant CO and constant turbulent velocity
models after 1 Myr. Observations that yield peak-to-trough ratios higher
than those produced by a disk with constant CO abundance and zero
microturbulence might indirectly indicate CO chemical depletion.

\begin{figure*}[ht]
\centering
\begin{tabular}{@{}cc@{}}
\includegraphics[width=0.45\textwidth]{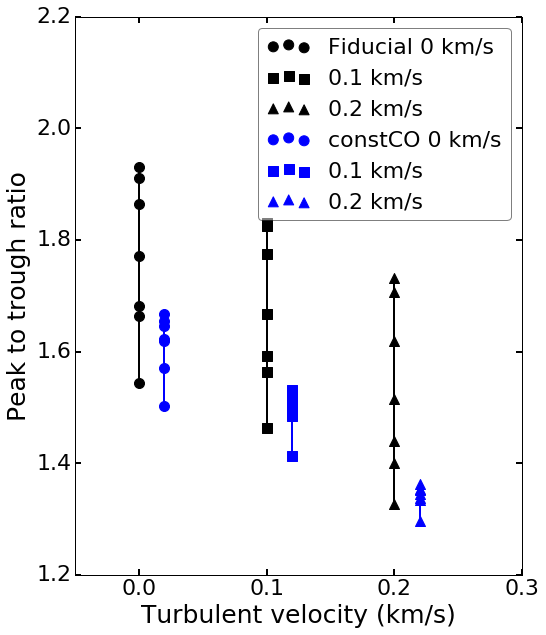} &
\includegraphics[width=0.45\textwidth]{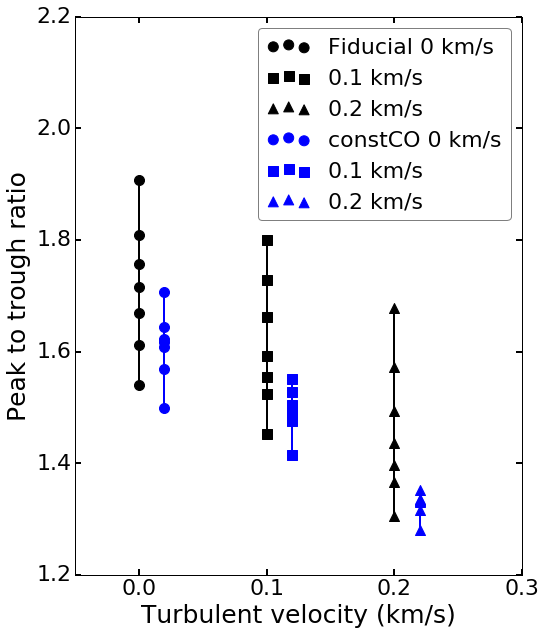} \\
 \end{tabular}
\caption{The peak-to-trough ratios of the CO $\jj32$ line for
the fiducial and constant CO models. The constant CO models are
displaced slightly to the right in each case for clarity. The
$0.015 \Msun$ disk is plotted on the left and the $0.03 \Msun$
model is on the right. For each disk, we include turbulent
speeds of 
$0, 0.1$ and $0.2 \kms$. Disk evolution always increases
peak-to-trough ratio, from a
combination of cooling and CO depletion for the fiducial models
and cooling only for the constant-CO models. Models with the
same turbulent velocity can produce a wide range of
peak-to-trough ratios, depending on disk age. Similarly, among disks with constant CO abundance, an old disk with $v_{tur} = 0.1 \kms$ produces the same peak-to-trough ratio as a young disk with no turbulence.}
\label{fig: p2t_tur}
\end{figure*}

We further illustrate the uncertainties in estimating the turbulent
velocity using peak-to-trough ratios by plotting the peak-to-trough
ratios from different models against the turbulent velocities used to
set up each model. The peak-to-trough ratios of the CO $\jj32$ line from
the fiducial model (with CO chemical depletion) and constant CO models
with turbulent velocities of $0, 0.1$ and $0.2 \kms$ are shown in Figure
\ref{fig: p2t_tur}. { For each model/velocity pair, the
peak-to-trough ratio increases over time, creating a vertical
spread of points.} Models with the same turbulent
velocity can produce a wide range of peak-to-trough ratios as CO
depletion increases the peak-to-trough ratio over time. { One
can easily conclude zero turbulence from a disk with CO
depletion and a true turbulent velocity of $0.1 \kms$ if
assuming a constant CO abundance.} By assuming a
constant CO/H$_2 = 10^{-4}$ abundance ratio { in radius and time}, one can easily conclude that there is
no turbulence in a CO-depleted disk with an RMS turbulent velocity of
$0.2 \kms$. For example, a $1$ Myr old with constant CO abundance and no
turbulence has the same peak-to-trough ratio as a $2$ Myr disk with
chemical CO depletion (fiducial model) and a turbulent velocity of $0.2
\kms$.



\begin{figure*}[ht]
\centering
\begin{tabular}{@{}cc@{}}
\includegraphics[width=0.45\textwidth]{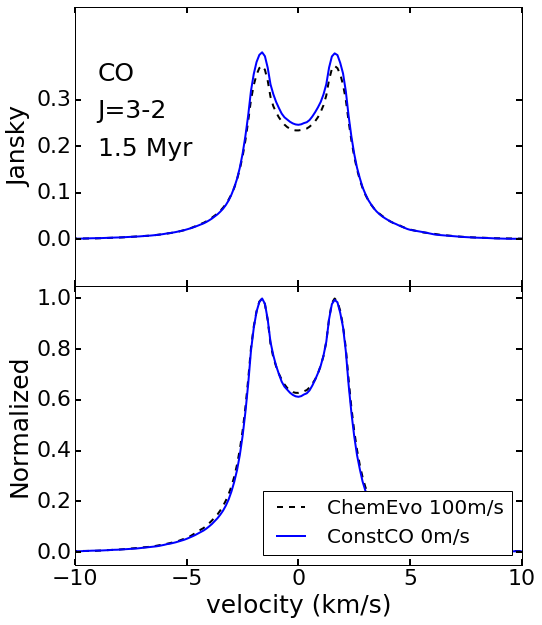} &
\includegraphics[width=0.45\textwidth]{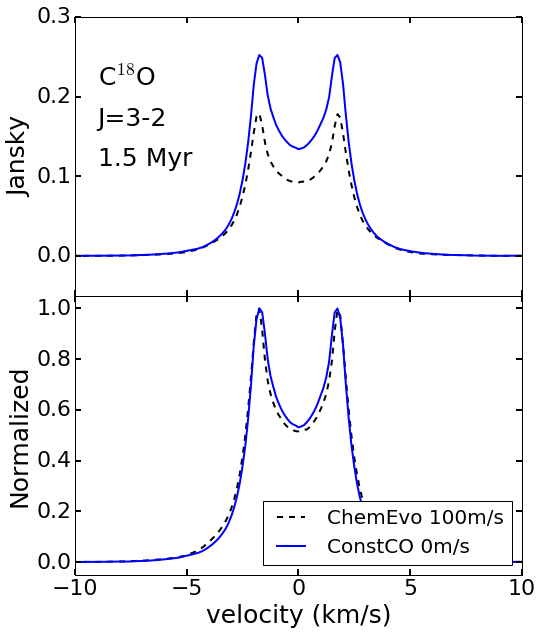} \\
 \end{tabular}
\caption{Comparisons of the line profiles of the CO $\jj32$ line from the $0.03 \Msun$ disk at $1.5$ Myr. CO lines are shown on the left and C$^{18}$O lines are shown on the right. We can see the line profiles of CO from the two models are very similar, resulting in close peak-to-rough ratios. The line profiles of C$^{18}$O are slightly different, but the difference may not be resolvable in observations due to noise. }
\label{fig: p03_CO_profiles}
\end{figure*}

{ From \S \ref{subsec: profiles_co_depletion} and \ref{subsec:
profiles_turbulent}, we see that CO depletion tends to produce wide
emission lines with large velocity spacing between the peaks. Is there a
guaranteed way to break the degeneracy between different models that
produce the same peak-to-trough ratio? We experimented with two methods
of disentangling chemical depletion and low turbulent speed: comparing
CO and \cooo\ line profiles, as advocated in Paper 2, and viewing CO
channel maps. From Figure \ref{fig: peak2trough}, we selected the $0.03
M_{\sun}$ disk at age 1.5~Myr---a stage when the fiducial model with
$v_{\rm tur} = 0.1 \kms$ and the constant-CO model with no turbulence
produce CO \jj32 emission with nearly the same peak-to-trough ratio.
Figure \ref{fig: p03_CO_profiles} shows the CO and \cooo\ line profiles
of both models. Although the CO line profiles are nearly identical, the
fiducial model has a slightly wider \cooo\ line at 10\% of maximum
intensity than the constant-CO model. However, the differences between
the two line profiles are very subtle, and are not likely to be
observable except at very high signal-to-noise.}

\begin{figure*}[ht]
\centering
\begin{tabular}{c}
\includegraphics[width=0.6\textwidth]{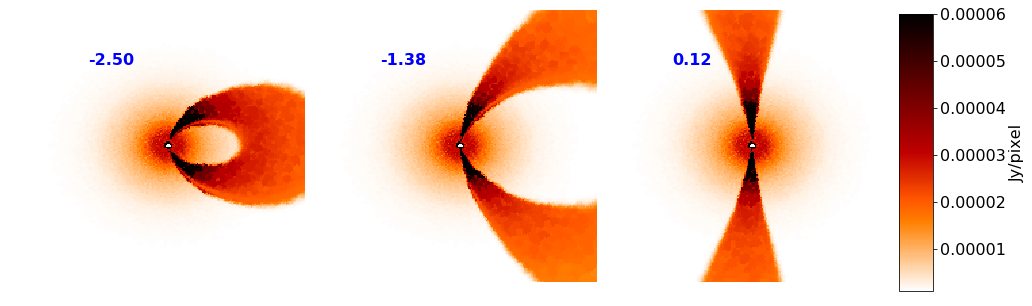}
\\
\includegraphics[width=0.6\textwidth]{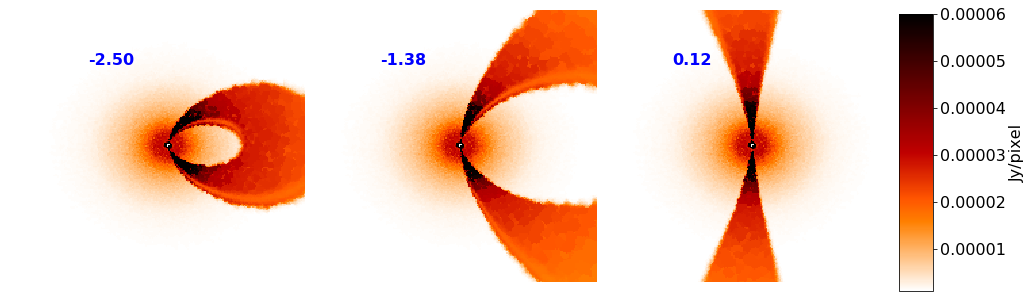}
\\
\includegraphics[width=0.6\textwidth]{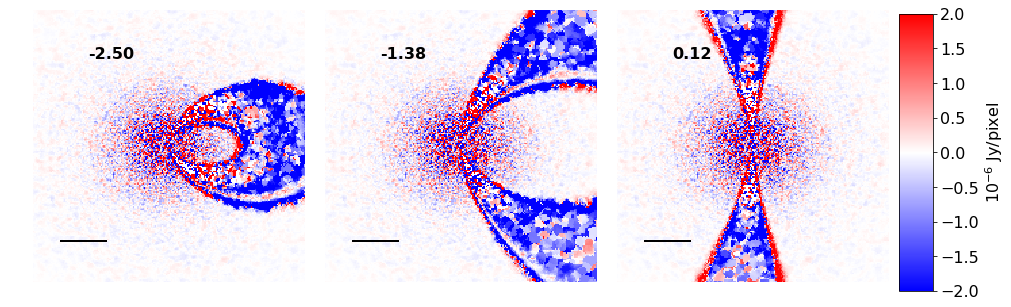}
\end{tabular}
\caption{{ Top:} CO \jj32 channel maps at several velocities from the
fiducial model with $v_{\rm tur} = 0.1 \kms$, age 1.5~Myr. { Middle:}
CO \jj32 channel map from the constant-CO model with no turbulence, age
1.5~Myr. { Bottom:} Difference image, (fiducial model - constant-CO
model) The diverging color map shows positive values in red and negative
values in blue. Each sub-panel is marked with the channel velocity in
\kms. Each channel map shows the same 60~AU$^2$ spatial domain. The
scalebars on the difference images represent $0.1 \arcsec = 14$~AU at
our assumed distance of 140~pc.}
\label{fig:chanmaps}
\end{figure*}

{ Figure \ref{fig:chanmaps} shows selected CO \jj32 channel maps of
the fiducial model with 0.1~\kms\ turbulence (top) and the constant-CO
model with no turbulence (middle), both with age 1.5~Myr. The bottom
panel shows the difference image, (fiducial - constant CO) in each
velocity window. Each panel is $60 \times 60$~AU$^2$. The most obvious
feature of the difference image is that the constant-CO model is
slightly brighter at larger radii, as positive values are shown in red
and negative values are rendered in blue. The fiducial model has
slightly more flux in the inner $\sim10$~AU of the disk. The differences
in any pixel are small, about $2 \times 10^{-6}$ Jy per
$0.003\times0.003 \arcsec$~pixel, but could add up to 20~mJy in a $0.3
\arcsec$~beam. Figure \ref{fig:chanmaps} suggests that the full 3-D data
cube could, in principle, constrain both the spatial distribution and
the velocity structure of the CO gas.}

\subsection{Effects of inclination}

So far we have set a $30$ degree inclination for our model disk, while
\citet{Simon_2015} assumed a $44$ degree inclination. The peak-to-trough
ratios presented here are therefore smaller than values presented in
\citet{Simon_2015}. To directly compare our model results with the
predictions of \citet{Simon_2015}, we compute the peak-to-trough ratios
of CO $\jj32$ for the $0.015$ \msun\ disk at a $44$ degree inclination,
assuming a constant CO abundance and zero turbulent velocities. We
choose to compare models with zero turbulent velocities and a constant
CO abundance to minimize the differences between this experiment and the
low turbulent velocity models in \citet{Simon_2015}. The peak-to-trough
ratios are plotted as the red lines in Fig. \ref{fig:
peak2trough_inclination}. Comparing the blue and red lines in the
figure, we can see that changing the inclination from $30$ to $44$
degrees increases the peak-to-trough ratios by about $0.3$. Translating
the peak-to-trough ratio into meaningful information about the turbulent
velocity requires detailed knowledge of the disk inclination. { For
spatially resolved disks, the inclination can be measured directly from
the major/minor axis ratio. We caution against using the peak-to-trough
ratio as a turbulence diagnostic for any disk without spatially resolved
data, where inclination is instead measured either from the SED or from
fitting the line profile.}

\begin{figure*}[ht]
\centering
\begin{tabular}{@{}cc@{}}
\includegraphics[width=0.45\textwidth]{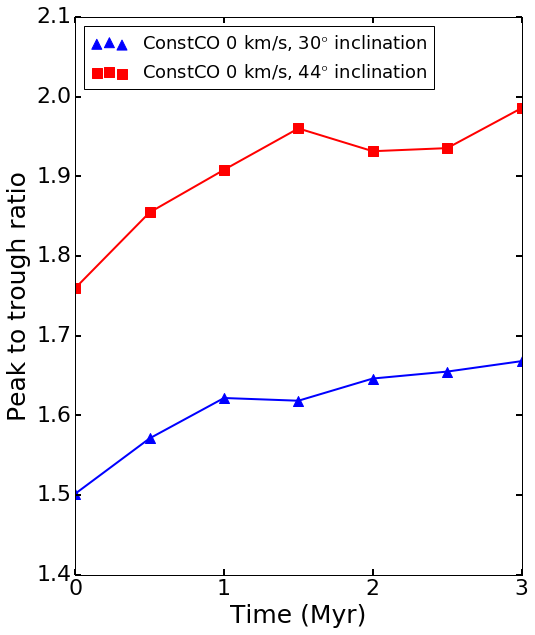} &
\includegraphics[width=0.45\textwidth]{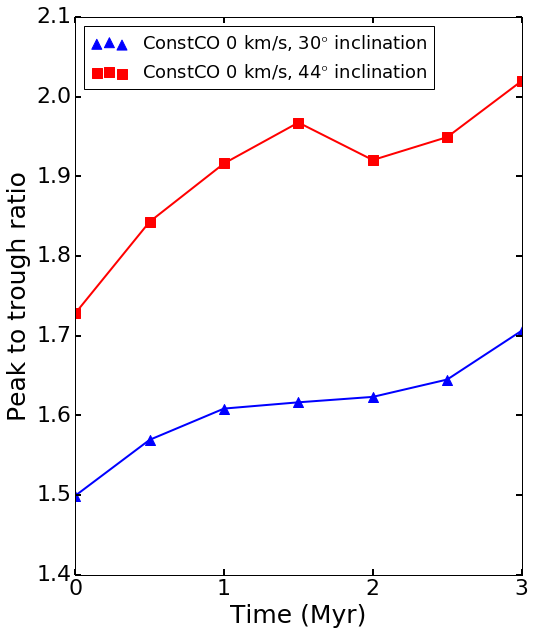} \\
 \end{tabular}
\caption{Comparison of the peak-to-trough ratios of the CO $\jj32$ line with 30 and 44 degree inclinations. The $0.015 \Msun$ disk is plotted on the left and the $0.03 \Msun$ model is on the right. To avoid the complications of other factors, we compare the models with a constant CO abundance and zero turbulent velocities. By changing the inclination from $30$ to $44$ degrees, the peak-to-trough ratios increase about $0.3$. }
\label{fig: peak2trough_inclination}
\end{figure*}

\section{Conclusion}
\label{sec:conclusion}

We demonstrate how the thermal and chemical evolution of a
protoplanetary disk complicates the measurement of the turbulent
velocity---a fundamental disk property that controls the first stages of
planet formation as grains assemble into pebbles. We show that the
peak-to-trough ratio could vary by up to $25\%$ due to the disk
evolution over time, as the CO abundance distribution changes and the
disk cools. One would underestimate the RMS turbulent speed by as much
as $200 \ms$ by assuming a constant CO/H$_2$ abundance ratio { of
$10^{-4}$} in a disk with chemical CO depletion. Even when chemical
depletion is not operating, simple disk cooling as the star evolves can
give an old disk with $v_{tur} = 0.1 \kms$ and a young disk with no
turbulence the same observed peak-to-trough ratio.

Quick observations of the abundant $^{12}$C$^{16}$O molecule are, by
themselves, inadequate diagnostics of disk turbulence: more detailed
information about the disk temperature and CO abundance distribution is
required. Deeper observations of multiple isotopologues { that yield
high signal-to-noise line profiles can help constrain the CO abundance
distribution, as can spatially resolved data (Paper 2, \S
\ref{sec:profile})}. In disks where the normalized line profiles are
substantially wider for the rare isotopologues C$^{17}$O and C$^{18}$O
than for CO and $^{13}$CO, chemical depletion is well underway and the
CO/H$_2$ abundance ratio is not constant with radius. { Considering
the uncertainties in CO abundance distributions pointed out here and the
difficulty of measuring disk temperatures from one or two emission lines
only \citep[e.g.][]{guilloteau12,Teague_2016}, care should be taken when
measuring turbulent speed based on a single peak-to-trough ratio.}



Work by MY, KW, SDR and NJT was supported by NASA grant NNX10AH28G
and further work by MY and SDR was supported by NSF grant 1055910.
This work was performed in part at the Jet Propulsion Laboratory, California
Institute of Technology. NJT was supported by grant 13-OSS13-0114 from
the NASA Origins of Solar Systems program. 
MY was supported by a Continuing Fellowship from the University of
Texas at Austin. We acknowledge helpful input from J. Simon, E.
Bergin and J. Lacy. { We thank the referee for comments that
improved the quality of the paper.}

\software{RADMC (http://www.ita.uni-heidelberg.de/ dullemond/software/radmc-3d/), LIME \citep{Brinch_Hogerheijde_LIME_2010}}

\addcontentsline{toc}{chapter}{Bibliography} 
\bibliographystyle{aasjournal}
\bibliography{references}

\end{document}